\documentclass[manuscript]{aastex}
\usepackage[usenames]{color}

\shorttitle{The Magnetic Structure of Solar Prominence Cavities}
\shortauthors{B\c ak-St\c e\' slicka et al.}

\begin{document}


\title{The Magnetic Structure of Solar Prominence Cavities: New Observational Signature Revealed by Coronal Magnetometry}

\author{Urszula B\c ak-St\c e\' slicka\altaffilmark{1,2}, Sarah E. Gibson\altaffilmark{3}, Yuhong Fan\altaffilmark{3}, Christian Bethge\altaffilmark{3}, Blake Forland\altaffilmark{4}, Laurel A. Rachmeler\altaffilmark{5}}

\altaffiltext{1}{Astronomical Institute, University of Wroc{\l}aw, ul. Kopernika 11, 51-622 Wroc{\l}aw, Poland}
\altaffiltext{2}{Visiting Scientist, High Altitude Observatory, NCAR, P.O. Box 3000, Boulder, CO 80307, USA}
\altaffiltext{3}{High Altitude Observatory, NCAR, P.O. Box 3000, Boulder, CO 80307, USA}
\altaffiltext{4}{Metropolitan State College of Denver, P.O. Box 173362, Denver, CO 80217-3362, USA}
\altaffiltext{5}{Royal Observatory of Belgium, Avenue Circulaire 3, 1180 Brussels, Belgium}
\begin{abstract}

The Coronal Multi-channel Polarimeter (CoMP) obtains daily full-Sun above-the-limb coronal observations in linear polarization, allowing for the first time a diagnostic of the coronal magnetic field direction in quiescent prominence cavities. We find that these cavities consistently possess a characteristic ``lagomorphic'' signature in linear polarization indicating twist or shear extending up into the cavity above the neutral line. We demonstrate that such a signature may be explained by a magnetic flux-rope model, a topology with implications for solar eruptions. We find corroborating evidence for a flux rope structure in the pattern of concentric rings within cavities seen in CoMP line-of-sight velocity. 

\end{abstract}



\section{Introduction}

Dark cavities are often part of coronal mass ejections (CMEs), surrounding bright cores identified as erupting prominences \citep{illing1986}. Non-erupting, or quiescent cavities also exist in equilibrium and may be long-lived. Understanding the magnetic structure of those cavities is important for understanding pre-CME configurations. Cavities are dark, elongated, elliptical structures with rarefied density \citep{fuller2009,gibson2010}. They have been observed in a wide wavelength range: mostly in white light \citep{gibson2006}, but also in radio, EUV and SXR \citep{marque2002, marque2004, hudson1999, hudson2000, heinzel2008, berger2012, reeves2012}. Cavities often surround quiescent prominences, especially in the polar crown regions \citep{tandberg1995}. They are long-lived, their structure changes slowly with time but they can also erupt as a CME \citep{maricic2004, vrsnak2004, gibson2006, regnier2011}. Cavities have been modeled as a flux rope \citep{low1994, low1995}, although the physical nature of cavities is still a subject of open research. Establishing their magnetic topology is important for choosing between models for CME eruptive drivers.

Measurements of the magnetic field in the solar corona are not trivial (\citet{lin2004}, and references therein). \cite{firor1962} showed that infrared forbidden lines of Fe XIII may be used to determine physical properties of coronal plasma. \cite{charvin1965} showed that direction of the magnetic field in the plane-of-sky (POS) can be determined using linear polarization signals from these forbidden coronal lines. The new Coronal Multi-Channel Polarimeter (CoMP), recently installed at the Mauna Loa Solar Observatory (MLSO) in Hawaii, makes daily observations of the lower corona with a field of view of about $1.04$ to $1.4$ solar radii. Since October 2010, CoMP has measured the magnetic field in the solar corona via the polarimetric signal (Stokes I, Q, U, V) of the forbidden lines of Fe XIII at 1074.7 nm and 1079.8 nm \citep{tomczyk2008}. The circular polarization (Stokes V) gives us information about the strength of the magnetic field along the line-of-sight (LOS). Due to the very low intensity of the circular polarization signal, long integration times on the order of hours are required. Linear polarization has a much stronger signal and constrains
the direction of the magnetic field in the plane-of-sky (POS) (see discussion below). CoMP also measures the LOS plasma velocity from observations at different wavelengths.

Early, prototype CoMP observations of cavities led to interesting results. \cite{schmit2009} analyzed observations from short CoMP observing runs at the National Solar Observatory in 2005 and found, for the first time, Doppler velocities of $5–-10$ km s$^{-1}$ within a coronal cavity. \citet{dove2011} found that the same observations showed that the cavity's signature in linear polarization was consistent with a spheromak-type magnetic flux rope model \citep{gibson1998}.
The \cite{dove2011} study analyzed only one cavity, however, motivating us to perform a more comprehensive study of cavity signatures in linear polarization, making use of the daily CoMP observations now available. We interpret the observations in a similar manner to that of \cite{dove2011}, applying the FORWARD codes\footnote{http://people.hao.ucar.edu/sgibson/FORWARD/} to an MHD model to yield synthesized CoMP observables \citep{judge2001}. We find however that an MHD model of a arched cylindrical flux rope \citep{fan2010} is a better fit than a spheromak to model the CoMP cavities that we survey, most of which surround polar crown filaments.

\section{{\bf CoMP Results}}
\begin{figure}[t]
\includegraphics[scale=0.77]{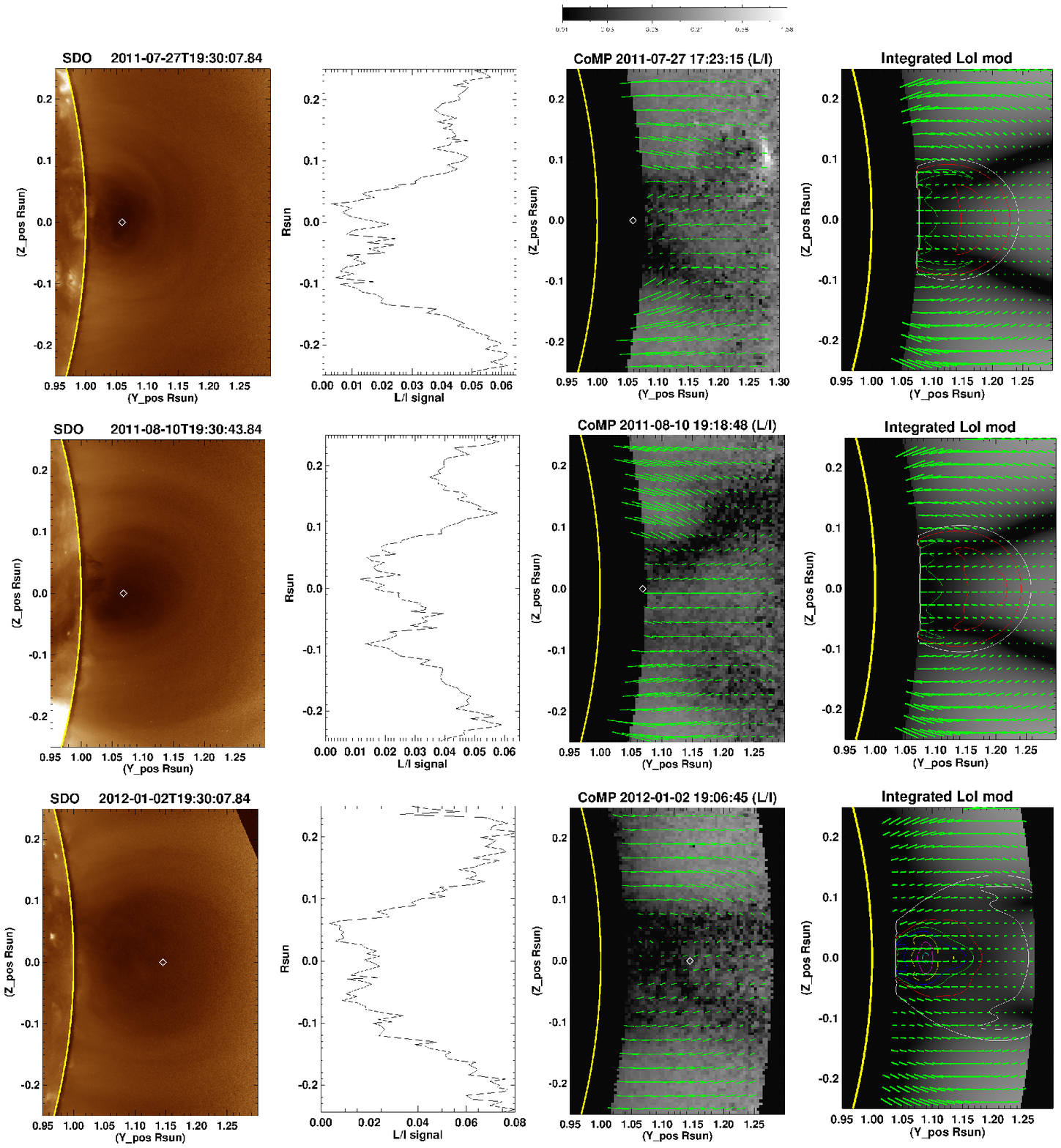}
\caption{{\em First column}: Examples of cavities observed by SDO/AIA $193$ \AA. {\em Second column}: L/I profile across polar-angle cuts at the cavity center height ($1.07$, $1.08$ and $1.14$ R$_\odot$ from top to bottom). {\em Third column}  LOS-integrated L/I for CoMP observations of three cavities. Direction of Stokes linear polarization vectors (integrated through the LOS) is shown as green lines. The edge of the solar disk is indicated by the curved yellow lines. The occulting disk of CoMP extends to $1.05$ R$\odot$. {\em Fourth column}: LOS-integrated Stokes L/I  for forward-calculated 3d flux rope model, where the the apex height of the flux rope axis matches the height of the center of the cavity observation calculated with AIA images (white diamonds). Contours show the current density of the simulated flux rope.
\label{fig1}}
\end{figure}

\begin{figure}
\includegraphics[scale=1.5]{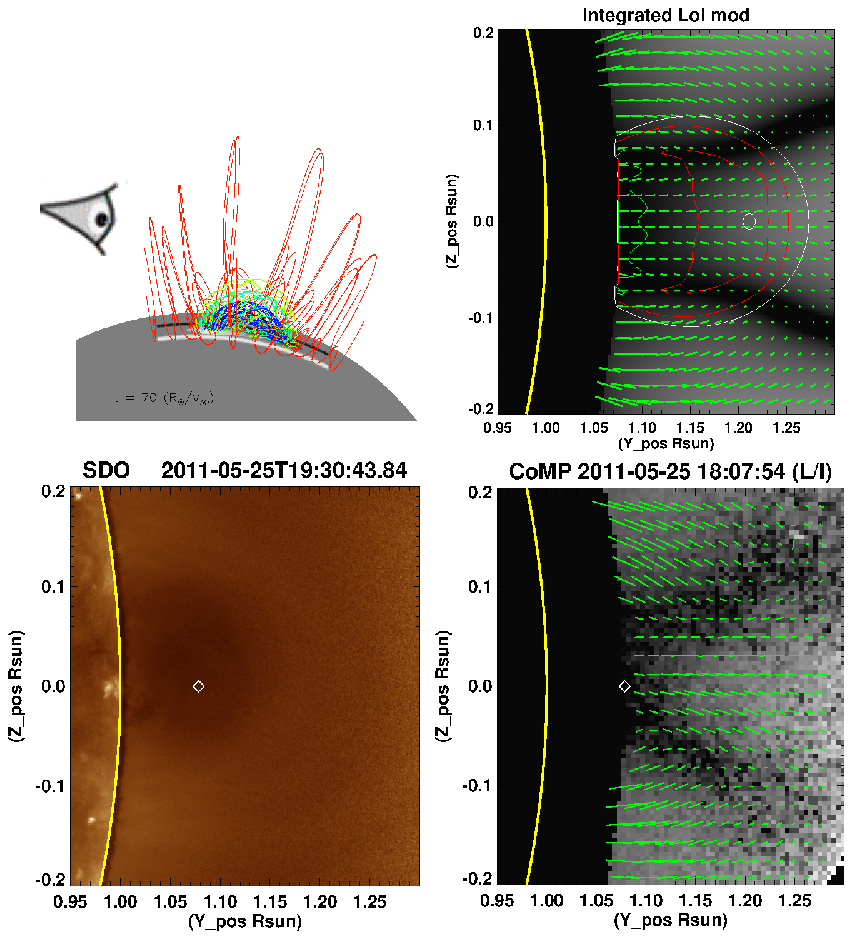}
\caption{{\em Top row}: Flux rope field lines (left) and model linear polarization integrated along line of sight (right). {\em Bottom row}: Cavity observed by SDO/AIA 193 \AA$ $ (left) and CoMP observed linear polarization (right). White diamonds show the center of the cavity seen on AIA images, and contours simulation current density. \label{fig2}}
\end{figure}

We have surveyed daily images from the \textit{Solar Dynamics Observatory}/Atmospheric Imaging Assembly (\textit{SDO}/AIA 193\AA) for polar-crown cavities, and subsequently examined CoMP data (averaged over tens of minutes to hours to improve signal-to-noise ratio) to establish cavity signatures in linear polarization. We found a consistent pattern of a dark V or U shape above a central core in CoMP linear polarization in the location of observed AIA cavities (Figures \ref{fig1}-\ref{fig2}). The quiescent cavities that we studied were tunnel-like in morphology, with a longitudinal extension that often allowed them to remain visible for many days. We found that the CoMP signature was generally apparent throughout this time period. Overall, in observations during 78 days between May 2011 and December 2012, we found 68 different cavities with this characteristic linear-polarization structure that we term lagomorphic, due to its resemblance to a rabbit-head seen in silhouette.

To establish the significance of the lagomorph structure, we examined polar-angle cuts in L/I (degree of linear polarization) (see e.g., Figure \ref{fig1}).  We found the L/I inside the lagomorph core ranged from a few percent to tens of percent (depending on radial height) lower than the signal outside the cavity at the same height.  In all cases this represented a greater than $3 \sigma$ depletion.

\begin{figure}
\includegraphics[scale=1.5]{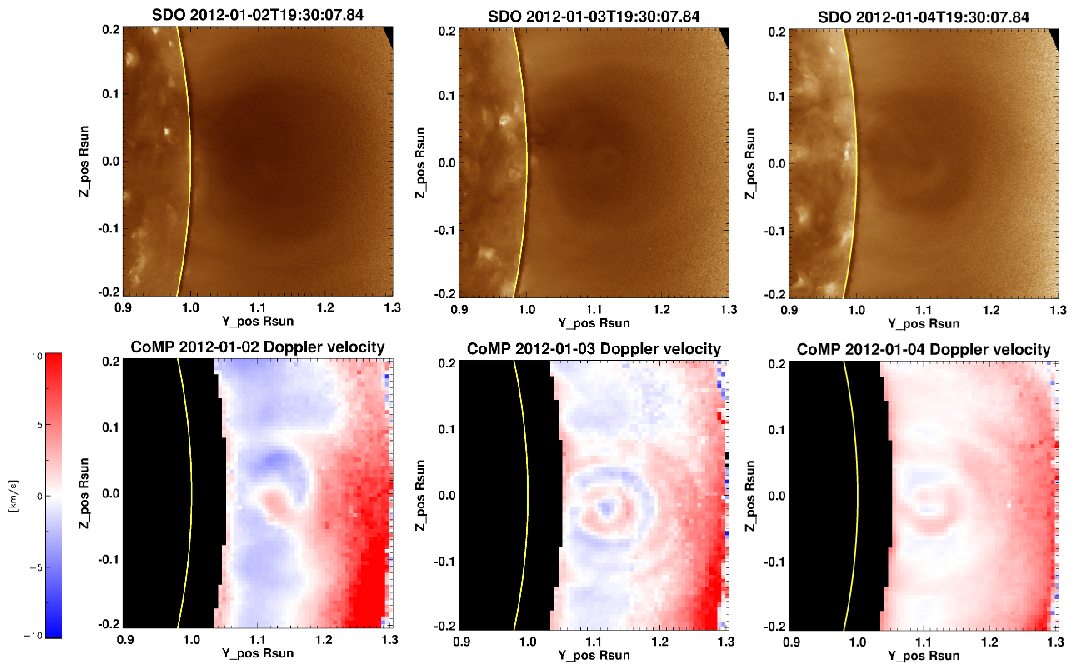}
\caption{{\em Top row}: Cavities observed by SDO/AIA $193$ \AA$ $ on 2012 January 2, January 3 and January 4. {\em Bottom row}: Doppler velocity from CoMP observations for the same cavities: these data can be downloaded from the MLSO web page and have been already corrected for East-West trend.\label{fig3}}
\end{figure}

Another interesting observation is found in the CoMP Doppler velocities. Our inspection of CoMP-observed cavities indicates that large-scale LOS flows as found by \cite{schmit2009} are common in cavities, but moreover that an interesting `bulls-eye' pattern may appear, with concentric circles of distinct values of flow along the LOS (Figure \ref{fig3}). Such a bulls-eye pattern is most easily observed in bigger cavities whose center is well above the CoMP coronagraph occulter, so that there is often no clear V-shape structure above the dark core because of the limited CoMP FOV. Bulls-eye flows in cavities have been observed to last for multiple days (e.g. Figure \ref{fig3}).

\subsection{{\bf Forward modeled flux rope}}
To interpret the new CoMP observations we have used the isothermal MHD model described in \cite{fan2010} with the temperature set to $T=1.5$ MK. In this three-dimensional MHD simulation, a twisted magnetic flux rope emerges into a pre-existing coronal potential arcade field. After the flux rope emergence is stopped, a quasi-static rise of the flux rope is observed. When the slow rise reaches a critical height, the flux rope accelerates and is rapidly ejected \citep{fan2010}. Figure \ref{fig2} shows an example, for one pre-eruption time step, of magnetic field lines and forward modeled linear polarization from this simulation. 

The magnitude of linear polarization depends on the angle $\theta$ between the direction of the local magnetic field and the LOS ($L\propto sin^{2}$ $\theta$, where L=$\sqrt{Q^2+U^2}$ is the total linear polarization). The strongest signal in linear polarization occurs when  the magnetic field is in the plane of sky ($\theta=90^{\circ}$). In the interpretation of such observations, examining the nulls in L is very useful (\cite{rachmeler2012} and references therein). Linear polarization goes to $0$ when $\theta=0^{\circ},180^{\circ}$ and the magnetic field is aligned with the LOS. The signal can also become unpolarized due to the Van Vleck effect ($L=0$ when the angle between the direction of the local magnetic field and the local vertical, $\vartheta$, is equal to $\vartheta_{VV}=54.7^{\circ}$). The Van Vleck effect also changes the direction of the linear polarization. If $\vartheta<\vartheta_{VV}$ then the direction of the linear polarization is parallel to the direction of the magnetic field in the POS. If $\vartheta>\vartheta_{VV}$, the direction of the linear polarization is perpendicular to the direction of the magnetic field in the POS. Consequently, vectors of linear polarization tend to be radial \citep{arnaud1987}, but it is possible that the $90^{\circ}$ ambiguity can be removed if the locations of $\vartheta=\vartheta_{VV}$ can be identified.

The forward-modeled linear polarization for the flux ropes shown in Figure \ref{fig1} (right column) show lagomorphs similar to that observed by CoMP.
The dark core in the middle of the cavity is due to the flux rope axis being oriented along the LOS, combined with Van Vleck inversions in the lower part of the flux rope. The V or U-shape (depending on height of the flux rope axis) structure above the dark core are due to Van Vleck inversions in the surrounding arcade and top of the flux rope.  
Depending on flux-rope axis height, the angle and location of  the lagomorph ``ears'' changes.  A potential arcade field (not presented in Figure \ref{fig1}), without an underlying rope, would produce only ears (i.e., a V-shape structure) without a dark central core, and we see that conversely a high-axis flux rope might have ears lying above the CoMP field of view.  Overlaid contours show current density of the simulation, and indicate that the ears tilt outwards just above the outer boundary of the flux rope.  Figure \ref{fig1} shows a similar trend in CoMP observations, from V-shaped ears for the lower cavity centers (top and middle), to more block-like horizontal type structure (no ears) for the higher cavity center (bottom).  
In four observed cases (one of them is shown in Figure \ref{fig1}, middle row) a dark central core is not clearly seen. It is possible that this is due to a curvature effect: \cite{rachmeler2} forward modeled linear polarization of an axisymmetric model of a flux rope (encircling the Sun), which is less curved than the arched cylindrical flux rope model we used. They found that linear polarization showed a similar lagomorph structure but without a clear dark core. Thus, these cases might imply that the cavity is elongated along the LOS. These measurements motivate 
future work to clearly establish how size and morphology scale between cavity and lagomorph, and how twist and curvature may affect the degree of linear polarization.

\section{Conclusions}
The linear-polarization lagomorph indicates shear or twist above neutral lines, but the question remains as to whether the flux rope model is the only possible model to explain these cavity observations. \cite{rachmeler2} has compared forward-modeled linear polarization signals from a sheared arcade, an arched cylindrical flux rope model, and a spheromak flux rope and found that all three models have distinct polarization signatures, especially when the direction of the linear polarization vectors are taken into account. The lagomorph morphology of linear polarization may be consistent with both the sheared arcade and flux rope model, but differs from the ring-like morphology predicted by the spheromak model and as seen in the CoMP observations described in \cite{dove2011}. One potential distinguisher between sheared-arcade and flux-rope model is the direction of the linear polarization vectors, which, below the maximum shear axis appear radial for the flux-rope model and horizontal for the sheared-arcade model \citep{rachmeler2}. Unfortunately, only a few cavities were large enough to investigate this: one such is shown in the bottom row of Figure \ref{fig1} - and for this case the vectors are more consistent with the flux-rope model. Perhaps the strongest corroborating evidence for the flux rope model is the bulls-eyes pattern in Doppler velocity images, which  suggests  flows along flux surfaces of a magnetic flux rope. It is also significant that the lagomorph core extends upwards into the cavity, since it implies that the axial field is not concentrated solely at lower heights where the prominence lies.
We note that, depending on the prominence type and even time of solar cycle it is possible that not all cavities are topologically similar. For example, the 2005 spheromak cavity of \cite{dove2011} was not a longitudinally extended polar-crown prominence cavity like the cases examined in this paper. Further analysis of a broader range of prominences and cavities is needed to establish the circumstances under which a given magnetic topology is likely to exist. 

In summary, we have used new CoMP observations to search for prominence-cavity signatures in linear polarization. We interpreted these observations using forward modeling to calculate synthetic CoMP-like data. We have found 68 different cavities with characteristic lagomorph structures in linear polarization observed by \textit{SDO}/AIA during 78 days between May 2011 and December 2012. Such signatures are well-explained by a flux rope topology, as are observations of observed bulls-eye patterns in LOS velocity. We conclude that the arched cylindrical magnetic flux rope is an appropriate model for most polar crown prominence cavities. This magnetic topology should thus be taken into account in determining likely physical processes that may ultimately destabilize cavity equilibria.

\acknowledgments
We thank  Steven Tomczyk for internal review of this manuscript. The authors acknowledge helpful discussion with Don Schmit, Durgesh Tripathi, Jim Dove, Guliana de Toma, and B. C. Low. We thank Leonard Sitongia for helping us with the CoMP data. This work was supported by NASA LWS grant NNX09AJ89G. UBS also acknowledges financial support from the Polish National Science Centre grant 2011/03/B/ST9/00104 and Human Capital Programme grant financed by the European Social Fund. BF thanks the University of Colorado Research Experience for Undergraduates (REU) program and the High Altitude Observatory where he was an REU student in 2011. LAR acknowledges financial support from STFC (UK). The CoMP data was provided courtesy of the Mauna Loa Solar Observatory, operated by the High Altitude Observatory, as part of the National Center for Atmospheric Research (NCAR). NCAR is supported by the National Science Foundation.


\begin{thebibliography}{}
\bibitem[Arnaud \& Newkirk (1987)]{arnaud1987} Arnaud, J. \& Newkirk, Jr., G. 1987, \aap, 178, 263  
\bibitem[Berger et al.(2012)]{berger2012} Berger, T. E., Liu, W. \& Low, B. C.  2012, \apj, 758, 37 
\bibitem[Charvin(1965)]{charvin1965} Charvin, P. 1965, Annales d$'$Astrophysique, 28, 877
\bibitem[Dove et al.(2011)]{dove2011} Dove, J. B., Gibson, S. E., Rachmeler, L. A., Tomczyk, S. \& Judge, P. 2011, \apjl, 731, L1
\bibitem[Fan(2010)]{fan2010} Fan Y. 2010, \apj, 719, 728
\bibitem[Firor \& Zirin(1962)]{firor1962} Firor, J. W. \& Zirin, H.  1962, \apj, 135, 122
\bibitem[Fuller \& Gibson(2009)]{fuller2009} Fuller, J. \& Gibson, S. E.  2009, \apj, 700, 1205
\bibitem[Gibson \& Low(1998)]{gibson1998} Gibson, S. E. \& Low, B. C. 1998, \apj, 493, 460
\bibitem[Gibson et al.(2006)]{gibson2006} Gibson, S. E., Foster, D., Burkepile, J., de Toma, G. \& Stanger, A. 2006, \apj, 641, 590
\bibitem[Gibson et al.(2010)]{gibson2010} Gibson, S. E., Kucera, T. A., Rastawicki, D., Dove, J., de Toma, G. 2010, \apj, 724, 1133
\bibitem[Heinzel et al.(2008)]{heinzel2008} Heinzel, P., Schmieder, B., F\'{a}rnik, F., Schwartz, P., Labrosse, N. et al. 2008, \apj, 686, 1383 
\bibitem[Hudson et al.(1999)]{hudson1999} Hudson, H. S., Acton, L. W., Harvey, K. A., \& McKenzie, D. M. 1999, \apj, 513, 83 
\bibitem[Hudson \& Schwenn (2000)]{hudson2000} Hudson, H. S. \& Schwenn, R. 2000, Adv. Space Res., 25, 1859 
\bibitem[Illing \& Hundhausen(1986)]{illing1986} Illing, R. M., \& Hundhausen, J. R. 1986, J. Geophys. Res., 91, 10951 
\bibitem[Judge \& Casini (2001)]{judge2001} Judge, P. G. \& Casini, R. 2001, Adv. Space Res., 25, 1859 in Astronomical Society of the Pacific Conference Series,
Vol. 236, Advanced Solar Polarimetry – Theory, Observation, and Instrumentation, ed. M. Sigwarth, 503
\bibitem[Lin et al.(2004)]{lin2004} Lin, H., Kuhn, J. R. \& Coulter, R. 2004, \apj, 613, L177 
\bibitem[Low(1994)]{low1994} Low, B. C. 1994, Phys. Plasmas, 1, 1684
\bibitem[Low \& Hundhausen(1995)]{low1995} Low, B. C. \& Hundhausen, J. R. 1995, \apj, 443, 818     
\bibitem[Mari\v{c}i\v{c} et al.(2004)]{maricic2004} Mari\v{c}i\v{c}, D., Vr\v{s}nak, B., Stanger, A. L. \& Veronig, A. M., 2004, \solphys, 225, 337
\bibitem[Marqu\'{e}(2004)]{marque2004} Marqu\'{e}, C. 2004, \apj, 602, 1037   
\bibitem[Marqu\'{e} et al.(2002)]{marque2002} Marqu\'{e}, C., Lantos, P. \& Delaboudin\`{e}re, J.-P. 2004, \aap, 387, 317       
\bibitem[Rachmeler et al.(2012)]{rachmeler2012} Rachmeler, L. A., Casini, R. \& Gibson, S. 2012,  Astronomical Society of the Pacific Conference Series, Volume 463, eds.  Rimmele, Collados Vera, Tritschler, Woeger, Carlsson, Schlichenmaier, Cadavid, Berdyugina, Knoelker, Berger, Gilbert, Socas Navarro, Goode
\bibitem[Rachmeler et al.(2013)]{rachmeler2} Rachmeler, L. A., Gibson, S. E. \& Dove, J. 2013, submitted to \solphys
\bibitem[Reeves et al.(2012)]{reeves2012} Reeves, K. K., Gibson, S. E., Kucera, T. A., Hudson, H. S. \& Kano, R. 2010, \apj, 746, 146  
\bibitem[R\'{e}gnier et al.(2011)]{regnier2011} R\'{e}gnier, S., Walsh, R. W. \& Alexander, C. E. 2011, \aap, 533, L1  
\bibitem[Schmit et al.(2009)]{schmit2009} Schmit, D. J., Gibson, S. E., Tomczyk, S., Reeves, K. K., Sterling, C. et al. 2009, \apj, 700, L96  
\bibitem[Tandberg-Hanssen(1995)]{tandberg1995} Tandberg-Hanssen, E. 1995, The nature of solar prominences, 2nd edn. (Dordrecht: Kluwer)
\bibitem[Tomczyk et al.(2008)]{tomczyk2008} Tomczyk, S., Card, G. L., Darnell, T., Elmore, D. F., Lull, R. et al. 2008, \solphys, 247, 411
\bibitem[Vr\v{s}nak et al.(2004)]{vrsnak2004} Vr\v{s}nak, B., Mari\v{c}i\v{c}, D., Stanger, A. L. \& Veronig, A. M., 2004, \solphys, 225, 355

\end{thebibliography}
\end{document}